\documentclass{phbauth}        
\usepackage{graphicx}

\begin{document}

\begin{frontmatter}

\title{Antiferromagnetism in the 2D Hubbard Model -- Phase Transition and
  Local Quantities}

\author[address1]{Roland M\"unzner},
\author[address1]{Ferdinando Mancini\thanksref{thank1}},
\author[address1]{Adolfo Avella}

\address[address1]{Dipartimento di Scienze Fisiche "E.R. Caianiello"
  -- Unit\`a INFM di Salerno, Universit\`a di Salerno, Via S.\ Allende,
  84~081 Baronissi, Italy} 

\thanks[thank1]{Corresponding author. \\
  E-mail: mancini@vaxsa.csied.unisa.it.\\
  Fax: +39--089--965275}  

\begin{abstract}
  We present a first study of the antiferromagnetic state in the $2D$
  $U$--$t$--$t^{\prime}$ model at finite temperatures by the composite
  operator method, providing simultaneously a fully self--consistent
  treatment of the paramagnetic and the AF phase.
  Near half filling the critical value of the Coulomb repulsion as a
  function of $t^{\prime}$ and the temperature dependence of the
  magnetization and internal energy have been studied. 
\end{abstract}


\end{frontmatter}

\section{Introduction}
\label{sec:int}
An antiferromagnetic phase is observed in most \emph{cuprate} high
$T_{\mathrm{c}}$ superconductors. On the other hand antiferromagnetic
correlations are considered to play an important role in the mechanism
of pair formation in those materials. 

The normal phase of $2D$ Hubbard model has been studied intensively by
the composite operator method (COM), leading to good agreement with
numerical results and experimental data \cite{Man95}. A study of the
AF phase in the $2D$ Hubbard model by COM is therefore supposed to
give some insight in both, the nature of antiferromagnetism and the
pair forming mechanism.  

\section{Results}
\label{sec:tf}
We investigate the Hubbard Hamiltonian on a $2D$ square lattice with
additional next--nearest neighbor hopping $t^{\prime}$
\begin{eqnarray*}
  H &=& \sum\limits_{i, j , \sigma} \left ( t_{ij} +
    t^{\prime}_{ij} - \delta_{ij} \mu \right ) c_{\sigma}^{\dagger}(i)
    c_{\sigma}(j)\\
    &&+ U \sum\limits_{i} n_{\uparrow}(i)n_{\downarrow}(i)
\end{eqnarray*}
by means of COM \cite{Man95}, using the Hubbard operators
$\xi_{\sigma}(i)=c_{\sigma}(i)\left( 1-n_{\bar{\sigma}}(i)\right)$
and $\eta_{\sigma}(i)=c_{\sigma}(i)n_{\bar{\sigma}}(i)$ as
components of the basic spinor $\Psi(i)$.

To describe the AF phase we introduce a coarse grained, bipartite
lattice with lattice constant $a$, overlaying the chemical lattice by
grouping together pairs of nearest neighbors in the usual way. For the
two sublattices $A$ and $B$ we assume a symmetric electronic state,
$\left\langle A, \sigma\right\rangle = \left\langle B,
  \bar{\sigma} \right\rangle$.

The retarded Greens functions $G^{XY} \left(\mathbf{k},\omega\right) =
\left\langle \left \{ \Psi^{\dagger}(i), \Psi(j)\right \}
\right\rangle_{\mathrm{F.T.}}$ with $X,Y \in \left \{A,B\right\}$
in the static approximation where finite life time effects are
neglected are determined in a fully self--consistent way by the
constraints 
$\langle\xi_{\sigma}(i)\eta_{\sigma}^{\dagger}(i)\rangle=0$ and
$\langle\xi_{\uparrow}(i)\xi_{\uparrow}^{\dagger}(i)\rangle=
\langle\xi_{\downarrow}(i)\xi_{\downarrow}^{\dagger}(i)\rangle$,
emerging from the algebra of the Hubbard operators, namely from the
Pauli principle.

\begin{figure}[tbp]
  \begin{center}\leavevmode
    \includegraphics[width=1.0\linewidth]{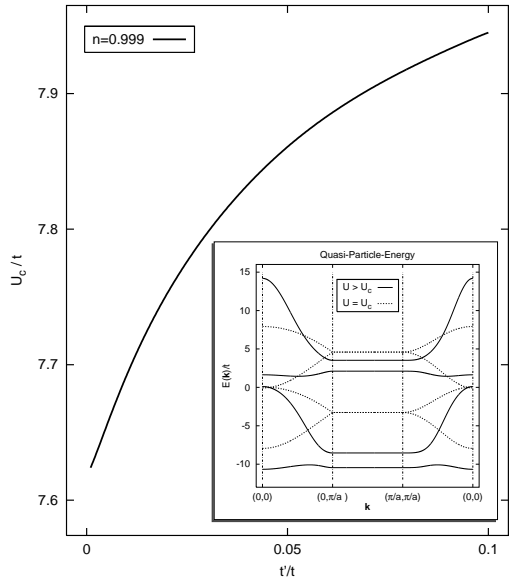}
    \caption{ 
      Critical Coulomb interaction at which the sublattice
      magnetization vanishes as a function of the next--nearest
      neighbor hopping $t'$ at $k_{\mathrm{B}}T=0.01 t$ and
      $n=0.999$. As inset the band structure is shown for
      $U_{\mathrm c}$ and $U > U_{\mathrm c}$ at the same
      temperature and particle density. 
      }
    \label{fig:U_c}
  \end{center}
\end{figure}

The critical value of the Coulomb interaction is given in
Fig.~\ref{fig:U_c} as function of $t^{\prime}$. 
$U_{\mathrm{c}}$ turns out to be significantly larger then in results
obtained by renormalization group techniques \cite{Alv98a}\cite{Alv98b},
which might be due to the fact, that the latter is valid only in the
weak coupling limit, whereas COM provides a good approximation for
intermediate and strong coupling as well. The band structure 
shows the opening of a gap at the transition line between the
paramagnetic and the AF phase. 

\begin{figure}[tbp]
  \begin{center}\leavevmode
    \includegraphics[width=1.0\linewidth]{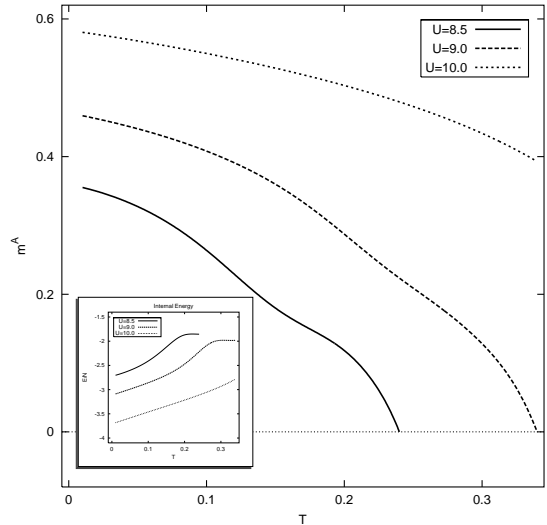}
    \caption{ 
      Sublattice magnetization as a function of temperature for
      various values of $U$ at $k_{\mathrm{B}}T=0.01 t$,
      $t^{\prime}=0.05t$ and $n=0.999$. As inset the corresponding
      values of the internal energy per site are plotted.  
      }\label{fig:m(T)}
  \end{center}
\end{figure}

In Fig.~\ref{fig:m(T)} the sublattice magnetization
$m=n^{A}_{\uparrow}(i)-n^{A}_{\downarrow}(i)$ is plotted versus
temperature for different values of the Coulomb interaction. With
decreasing magnetization the energy per site, which is not  sublattice
dependent because of the above symmetry assumption on the electronic
state, is increasing, reaching finally the paramagnetic value at
vanishing $m$. 

\section{Conclusions}
\label{sec:con}

First results for the AF phase of the $2D$ Hubbard model obtained by COM
within a fully self--consistent treatment were presented. Further
results and a detailed elaboration of the theoretical framework will
be presented elsewhere.



\end{document}